\documentstyle[epsfig]{aipproc}

\begin{document}
\title{A possible black hole binary in Mkn~501}

\author{Frank M. Rieger and Karl Mannheim}
\address{Universit\"ats-Sternwarte G\"ottingen\\ 
Geismarlandstr. 11, 37083 G\"ottingen, Germany}

\maketitle
\begin{abstract}
A simple binary model for Mkn~501 is considered under the assumption that the 
TeV and X-ray periodicity of $\sim 23$ days, observed during the flaring state 
in 1997, may be basically interpreted as a doppler-shifted flux modulation due 
to the orbital motion of the relativistic jet. 
For the typical jet properties inferred from emission models, we find an 
intrinsic orbital period of $(6-14)$ yrs and a centre-of-mass distance of 
$(2-3.5) \times 10^{16}$ cm, the binary thus being a very close 
system. If the binary separation corresponds to that at which 
gravitational radiation becomes dominant, one may obtain a maximum allowed
primary mass of $\sim 10^8\, M_{\odot}$ and a corresponding maximum allowed
secondary mass in the range of $\sim (1-9)\times 10^7\, M_{\odot}$ assuming
that gas accretion occurs on around the salpeter time scale.
Interestingly such values agree with the black hole masses expected from 
merger scenarios. 
\end{abstract}

\section*{Introduction}
\begin{figure}[htb]
       \vspace{0cm}
       \begin{center}
         \epsfxsize=8.0cm         
          \mbox{\epsffile{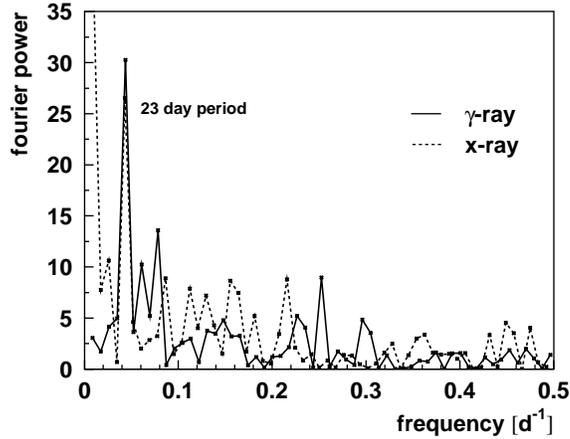}}
       \end{center}
       \vspace{-0.2cm} 
       \caption{Fourier analysis of the TeV and X-ray data from April to
        July 1997 using a normalized Lomb periodogram. The periodicity of
        $23$ days is clearly identifiable. Figure courtesy of D. Kranich.}
       \label{fourier}
   \end{figure}

Binary black hole systems (BBHS) are expected to be a common phenomenon 
in the universe. For, in hierarchical clustering scenarios, giant elliptical
galaxies (e.g. the host galaxy of Mkn~501) are thought to be products
of mergers between spiral galaxies. Since the brightest galaxies generally 
seem to contain massive black holes (BH) in their nuclei, a BBHS should be a 
natural consequence, e.g. \cite{begelman80}, \cite{merritt99}.
In the specific case of the gamma-ray blazar Mkn~501 at $z=0.034$, the 
existence of such a putative BBHS has been repeatedly invoked as a possible 
explanation of the observed misalignment between pc and kpc-jet 
\cite{conway95} or as a reason for the special evolution of its X-ray SED 
\cite{villata99}. 
The periodicity of $P_{\rm obs}\simeq 23$ days, recently observed in the 
X-ray and $\gamma$-ray band of Mkn~501 (\cite{kranich99}, \cite{nishikawa99};
see {\bf Fig.~\ref{fourier}}) may add another piece of evidence on such an 
internal configuration. 
Here we show, that this periodicity may be interpreted as being basically
caused by the orbital motion of the relativistic jet emerging from the
less massive BH in a close binary system \cite{rieger00}.
\section*{Modelling}
\begin{figure}[htb]
       \vspace{0cm}
       \begin{center}
         \epsfxsize=7cm         
          \mbox{\epsffile{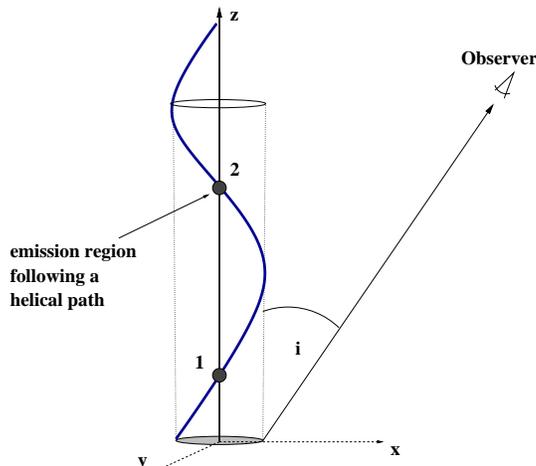}}
       \end{center}
       \vspace{-0.5cm} 
       \caption{Simple sketch of a helical jet path caused by the
        orbital motion in a BBHS.}
       \label{helix}
   \end{figure}

 Assuming that the periodicity observed in the flaring state in Mkn~501 is 
 associated with a relativistically moving feature in the jet (e.g. blob, 
 shock, knot) the flux modulation by Doppler boosting is given by the simple 
 formula
 \[S_{\nu}^{obs}=\delta(t)^{3+\alpha}\, S_{\nu}\,,\]
 where $\alpha$ is the spectral flux index. For a source region following
 a helical path (see {\bf Fig.~\ref{helix}}) the Doppler factor $\delta(t)$, 
 and thus the observed flux, is a periodical function of time and may be 
 written as
 \[\delta(t)=\frac{\sqrt{1-(v_z^2 + \Omega_k^2 R^2)/c^2}}{1-
    (v_z \cos i-\Omega_k R\,\sin i \sin \Omega_k\,t)/c}\,,\]
 where $\Omega_k$ is the keplerian orbital frequency, $R$ the centre-of-mass 
 distance, $i$ the jet inclination angle, $c$ the velocity of light and $v_z$
 the outflow velocity. 
 Due to the relativistic motion of the source region the observed period
 appears strongly shortened
 \[P_{\rm obs}=(1+z)\,(1-\frac{v_z}{c}\cos i)\, P_k\,.\]
 Thus, for an observed period of $P_{\rm obs}=23$ days, typical inclination 
 angles $i \simeq 1/\Gamma_b$ and bulk Lorentz factors $\Gamma_b=10-15$, e.g.
 \cite{mannheim96,spada99},  
 we have an intrinsic orbital period  $P_k = (6-14)\,\,{\rm yrs}\,$.
 Using the definition of the keplerian orbital frequency, one may also 
 derive an equation for the required binary mass ratio \cite{rieger00}
 \begin{eqnarray}
 \frac{M}{(m+M)^{2/3}} & = & \frac{P_{\rm obs}^{1/3}}{(2\,\pi\,[1+z]\,
                    G)^{1/3}}\,\frac{c}{\sin i}\,
                    \nonumber \\            
                    & &\times \frac{f^{1/(3+\alpha)}-1}{f^{1/(3+\alpha)}+1}\,
                        (1-\frac{v_z}{c}\,\cos i)^{2/3}\nonumber\,,
 \end{eqnarray} with $m$, $M$ the mass of the smaller/larger BH, respectively,
 $G$ the gravitational constant and $f \sim 8$ the observed TeV flux ratio 
 between maximum and minimum.\\  
 Applying an additional constraint by assuming that the current binary
 separation $d$ roughly corresponds to the separation at which gravitational
 radiation becomes dominant, we may set an upper limit on the allowed
 binary masses (note that for a more precise estimate the gravitational 
 time scale $\tau_{\rm grav}$ in \cite{rieger00} should be multiplied by a 
 factor of $4$ and accordingly, the rhs of Eq.~(10) by $0.25^{0.25}$, thus 
 enhancing the allowed upper limit for $m$ by roughly a factor of $3$). 
 These upper limits are calculated for the relevant inclination angles and 
 two different spectral indices in {\bf Tab.~\ref{tab}}, where $P_K$ denotes 
 the intrinsic orbital period, $\tau_{\rm l}=\tau_{\rm grav}/4$ the 
 gravitational lifetime and $P_p$ the period of gravitomagnetic jet precession.
 The calculation has been performed assuming a.) that the dynamical timescale 
 for (Eddington) gas accretion is of the order of the salpeter time scale, i.e 
 $t_{\rm gas} \sim t_{\rm s} \simeq 5 \times 10^7$ yrs (note that for 
 $t_{\rm gas} > t_{\rm s}$ the upper limit for $m$ can be reduced) and 
 b.) that the spectral index $\alpha=1.2$\,\,(in bracket: $\alpha =1.7$).
 The above mentioned constraint may be related to the key aspect that
 BL Lacs are less luminous radio galaxies with rather little signs of 
 cosmological evolution \cite{bade98}. They may thus correspond to 
 the final evolutionary stage of BBHSs with decreased mass accretion rates 
 \cite{villata99}. 
  \begin{table}[htb]
  \caption{Upper limit for the maximum binary masses.}
   \label{tab}
  \begin{tabular}{l@{\hspace{1cm}} l@{\hspace{1cm}}
                 l@{\hspace{1cm}}}
  $ i=1/\gamma_b$  & $1/10$ & $1/15$  \\ \tableline
  $ m\,[10^8\,M_{\odot}]$   & 0.13 \quad (0.18)   & 0.81 \quad (0.98) \\ 
  $ M\,[10^8\,M_{\odot}]$   & 1.11 \quad (0.92)   & 1.44 \quad (1.31) \\ 
  $ d\,[10^{16}\,\rm{cm}]$  & 2.58 \quad (2.48)   & 5.40 \quad (5.43)  \\ 
  $ P_k\,[\rm{yrs}]$        & 6.10                & 13.7    \\ 
  $ \tau_{\rm l}\,[10^7\,\rm{yrs}]$ 
                           & 1.57 \quad (1.30)   & 2.04 \quad (1.86) \\ 
  $ P_p\,[10^4\,\rm{yrs}]$  & 0.69 \quad (0.59)   & 5.88 \quad (6.05) \\ 
  \end{tabular} 
  \end{table}
 \vspace{0cm}
\section*{Conclusions}
 The results indicate that a BBHS in Mkn~501 may not be improbable
 since 
 1.) it might represent a natural explanation for the observed periodicity;
 2.) the putative binary black hole interpretation is supported by galaxy 
     merger evolution;
 3.) the derived masses are in line with such expectations from merger 
     scenarios, i.e. a secondary mass of $m \sim 10^7 \,M_\odot$ appear 
     to be in agreement with the concept that the galaxy swallowed in the 
     merger process was a (minor) spiral galaxy; 
 4.) its expected separation is similar to that suggested in a simple 
     unification scenario for BL Lacs and FSRQs \cite{villata99}\,;
 5.) its precessional period of $\sim 10^4$ yrs would agree with the 
     driving frequency found by Conway \& Wrobel \cite{conway95} in order to 
     explain the misalignment of the radio jet on pc and kpc-scale.\\
 We have also checked for consistency with several observations 
 (e.g. VLBI and X-ray (RXTE) data) and discussed the implications 
 elsewhere \cite{rieger00}. 
 To be sure, the presented model is rather simplified because small changes, 
 for example, in the maximum electron Lorentz factor or the magnetic field 
 along the trajectory of the emission region may further add to flux 
 variations.
 However, the general signature according to which a rotating jet in a BBHS
 may cause a periodical flux modulation remains robust.
 Generally, if the binary hypothesis is correct, the observable period in 
 Mkn~501 should remain similar during different outburst phases unless there 
 is evidence for a change in the general jet properties (i.e. in the bulk 
 Lorentz factor or inclination angle).

{\small Acknowledgement: K.M. acknowledges financial support from a 
Heisenberg-Fellowship and F.M.R. from DFG Ma 1545/2-2.}

\end{document}